\documentclass[a4paper,11pt]{article}
\pdfoutput=1 

\usepackage{jcappub} 
\usepackage{natbib}
\usepackage{microtype}
\usepackage[T1]{fontenc} 

\pdfoutput=1

\usepackage{graphicx}
\usepackage{bm}
\usepackage{float}
\usepackage{slashed}
\usepackage{color}

\usepackage[normalem]{ulem}
\usepackage{lipsum}
\usepackage[caption=false]{subfig}
\usepackage{multirow}

\hypersetup{
     colorlinks   = true,
     citecolor    = magenta,
     urlcolor     = blue,
     linkcolor    = magenta
}

\usepackage{listings}
\usepackage[dvipsnames]{xcolor}
\usepackage{orcidlink}

\title{Successful $\nu p$-process in neutrino-driven outflows in core-collapse supernovae}

\author[a]{Alexander Friedland,\,\orcidlink{0000-0002-5047-4680}}
\emailAdd{alexfr@slac.stanford.edu} 
\affiliation[a]{%
SLAC National Accelerator Laboratory,
Menlo Park, CA 94025, USA
}

\author[a,b,c]{Payel Mukhopadhyay\,\orcidlink{0000-0002-3954-2005}}
\emailAdd{pmukho@berkeley.edu} 
\affiliation[b]{Physics Department, Stanford University, Stanford, CA 94305, USA}
\affiliation[c]{Department of Physics, University of California, Berkeley, Berkeley, CA 94720, USA}

\author[a,d,e]{and Amol V.\ Patwardhan\,\orcidlink{0000-0002-2281-799X}}
\emailAdd{apatwa02@nyit.edu} 
\affiliation[d]{%
School of Physics and Astronomy, University of Minnesota, Minneapolis, MN 55455, USA
}
\affiliation[e]{%
Department of Physics, New York Institute of Technology, New York, New York 10023, USA
}

\date{\today}

\begin{document}


\abstract{The origin of the solar system abundances of several proton-rich isotopes, especially $^{92,94}$Mo and $^{96,98}$Ru, has been an enduring mystery in nuclear astrophysics.
An attractive proposal to solve this problem is the $\nu p$-process, which can operate in neutrino-driven outflows in a core-collapse supernova after the shock is launched. 
Years of detailed studies, however, have cast doubt over the ability of this process to generate sufficiently high absolute and relative amounts of various $p$-nuclei. The $\nu p$-process is also thought to be excluded by arguments based on the long-lived radionuclide $^{92}$Nb.
Here, we present explicit calculations, in which both the abundance ratios and the absolute yields of the $p$-nuclei up to $A\lesssim105$ are successfully reproduced, even when using the modern (medium enhanced) triple-$\alpha$ reaction rates. The process is also shown to produce the necessary amounts of $^{92}$Nb. The models are characterized by subsonic outflows and by the protoneutron star masses in the {$\gtrsim1.7 M_\odot$ range}. This suggests that the Mo and Ru $p$-nuclides observed in the Solar System were made in CCSN explosions characterized by an extended accretion stage.} 
\maketitle

\section{Introduction}

Most of the trans-iron elements in nature are synthesized by slow or rapid neutron capture, the $s$- and $r$-processes~\cite{Burbidge:1957vc,Kaeppeler:2010kk,Arnould:2007gh}. These processes, however, bypass 35 naturally occurring proton-rich isotopes, the origin of which has been a classic problem in nuclear astrophysics~\cite{Burbidge:1957vc,Cameron:1958vx,Rauscher:2013}.
%
Some of the $p$-nuclides can be created by the photo-disintegration of preexisting $s$-process isotopes in exploding stars (the $\gamma$-process)~\cite{Woosley:1978, Howard:1991, Travaglio:2011}. Yet, it was shown that the solar $s$-process isotope abundances are insufficient~\cite{Woosley:1978,ArnouldGoriely2003,Rauscher:2013} to explain all $p$-nuclei observations. Particularly challenging~\cite{ArnouldGoriely2003,Rauscher:2013} are $^{92,94}$Mo and $^{96,98}$Ru, whose abundances are comparable to those of the corresponding $s$-process isotopes~\cite{Woosley:1978,Lodders:2003}.  Even positing large $s$-process enrichment before a Type-Ia explosion, the observed abundances of $^{94}$Mo cannot be obtained~\cite{Travaglio:2014,Rauscher:2014fea}. 

One is thus led to nucleosynthetic paths involving proton capture. This requires not only a proton-rich medium, but also a specific temperature window, $1.5\,\mbox{GK}\lesssim T \lesssim 3$\,GK~\cite{Wanajo:2010mc}. In this window, the temperature is high enough to overcome Coulomb repulsion, but not so high as to bring heavy nuclei into quasi-equilibrium with the iron group~\cite{ClaytonBook}. The classical $rp$-process, however, encounters a difficulty: in typical astrophysical settings, the material spends less than a second in this temperature range, while the reaction chain includes several longer-lived isotopes, which create \emph{waiting points} that prevent the synthesis of heavier $p$-isotopes~\cite{Schatz:1998zz}.

An elegant proposal that evades this difficulty is the $\nu p$-process~\cite{Frohlich:2005ys,Pruet:2005qd,Wanajo:2006rp}. It  operates in 
the innermost ejecta of 
a core-collapse supernova (CCSN), in an outflow of high-entropy matter from the surface of the protoneutron star (PNS) that forms after the explosion is launched~\cite{Frohlich:2005ys,Pruet:2005qd,Wanajo:2006rp,Wanajo:2010mc,Arcones:2011zj,Eichler:2017,Nishimura:2019jlh,Rauscher:2019mcn, Fujibayashi:2015rma,Sasaki:2017jry, Xiong:2020ntn, Sasaki:2021ffa,Sasaki:2023ysp, Fisker:2009, Bliss:2014qiz, Bliss:2018djg, Jin:2020}. A physically plausible choice of neutrino and antineutrino spectra makes the outflow proton-rich (electron fraction $Y_e\sim 0.6$). As the outflow material expands and cools, some of the nucleons assemble into heavy elements. The large neutrino flux  not only powers the outflow, but also 
creates a subdominant neutron population in the proton-rich medium. These neutrons are promptly captured on the proton-rich seeds, in reactions that bypass the beta-decay waiting points.

Several quantitative studies, however, have identified a number of difficulties in explaining the observed $p$-nuclide abundances with the $\nu p$-process~\cite{Rauscher:2013, Fisker:2009, Bliss:2014qiz, Bliss:2018djg, Jin:2020}. Various calculations in these papers: (i) did not reproduce the observed solar system abundance ratios of $^{92}$Mo/$^{94}$Mo~\cite{Fisker:2009,Bliss:2014qiz,Bliss:2018djg} and $^{96}$Ru/$^{98}$Ru isotopes~\cite{Bliss:2018djg}; (ii) significantly underproduced the absolute amounts of $^{92,94}$Mo and $^{96,98}$Ru~\cite{Jin:2020} compared to the fiducial astrophysical model~\cite{Wanajo:2010mc,Woosley:1994ux}, especially when incorporating modern, medium-enhanced triple-$\alpha$ reaction rates~\cite{Beard:2017jpg}; (iii) significantly overproduced the lighter $p$-nuclides,   $^{74}$Se, $^{78}$Kr, $^{84}$Sr, compared to $^{92,94}$Mo and $^{96,98}$Ru~\cite{Bliss:2018djg}; (iv) did not account for~\cite{Rauscher:2013} the presence of the $^{92}$Nb isotope in the early solar system~\cite{Lugaro:2016zuf,Haba:2021PNAS}; and, finally, (v) observed that, under certain conditions, the final isotopic composition of the outflow could be driven to the neutron-rich side~\cite{Arcones:2011zj}. 

In view of such a long list of difficulties, it appeared unlikely that all these conditions could be satisfied within a consistent outflow model, without ad hoc adjustment of parameters to unphysical values.
As a result, the current understanding of the origin of the $p$-isotopes is considered to be as uncertain as ever~\cite{Rauscher:2013,Jin:2020}. Yet, as shown below, a subclass of CCSN explosions naturally hosts conditions in which all of these difficulties are resolved.

\section{Results} \label{sec:results}

\subsection{Snapshots of subsonic and supersonic outflows}

The $\nu p$-process involves out-of-equilibrium dynamics, which makes its yields very sensitive to the outflow hydrodynamics~\cite{Wanajo:2010mc}. To test the viability of the $\nu p$-mechanism, one needs to scan the range of possible outflow conditions. This cannot be achieved by computing yields for a specific high-fidelity numerical simulation: a negative result would not exclude the mechanism for all physically plausible explosion scenarios. Accordingly, we consider a suite of physical 1D models which span the range of conditions (the same argument is made in, e.g.,~\cite{Bliss:2018djg}). 
Our goal is to identify physical regimes in which the $\nu p$ process is viable. Once identified, these regimes can be targeted for follow-up studies with detailed multi-D simulations. 

\begin{figure}[tbh]
\centering
%
%
%
%
\includegraphics[width = 0.8\columnwidth]{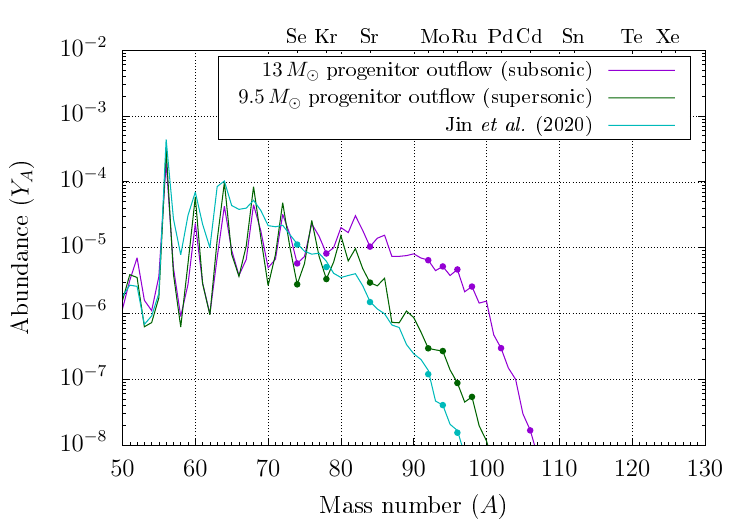}
\caption{Time snapshots of nucleosynthesis yields in our models with a subsonic outflow (13\,$M_\odot$ progenitor) and a supersonic outflow (9.5\,$M_\odot$ progenitor). Also reproduced is a calculation from Ref.~\cite{Jin:2020} with a parametrized supersonic outflow profile with entropy $S_\text{tot} = 80$. Nominal in-medium enhanced triple-$\alpha$ rates from Ref.~\cite{Jin:2020} are assumed throughout. The $p$-nuclei yields for $A > 90$ are greater by 1--2 orders of magnitude in the subsonic regime compared to the supersonic cases. 
}
\label{fig:Triple-alpha compared}
\end{figure}

{Several studies of the $\nu p$-process approach the problem by assuming a supersonic ansatz for the outflow. In this case, one can vary the entropy of the material and/or the position of the termination shock (e.g.,~\cite{Wanajo:2010mc,Jin:2020}).  
We, however, do not assume any fixed ansatz and instead vary \emph{the physical conditions} that control the outflow, namely, the properties of the PNS and the progenitor profile (see Appendix~\ref{subsec:outflow_calc} for details). An essential observation is that the nature of the outflow in a CCSN turns out to be sensitive to the details of these conditions~\cite{Friedland:2020ecy}: for sufficiently massive progenitors, the outflow stays subsonic. We find that these subsonic outflows \emph{generically} produce much greater yields of the desired $p$-nuclides than the supersonic ones.

This is shown in Fig.~\ref{fig:Triple-alpha compared}, where instantaneous yields are compared for our 9.5\,$M_\odot$ and  13\,$M_\odot$ progenitor models. In the first case, the outflow is in the traditional ``wind" regime~\cite{Duncan1986,Qian:1996xt,Thompson:2001}: it accelerates to supersonic speeds and runs into the surrounding slower moving material forming a termination shock. In the second case, the outflow speed remains subsonic throughout, thanks to the greater mass swept up by the front shock which creates a sufficient surrounding pressure.
In both cases, the outflows have $Y_e\sim0.6$ at small radii and the PNS radius is 19\,km, typical for $1\mbox{--}2$\,s after the shock launch. Notably, the PNS mass in the 13\,$M_\odot$ progenitor case is $1.8\,M_\odot$, so that total entropy per baryon $S_\text{tot} \simeq 90$ (or \textit{radiation} entropy per baryon $S\simeq74$), as discussed below. The details of both models are supplied in Table~\ref{table:nuparams} of the Appendix. }

For comparison, a calculation from Ref.~\cite{Jin:2020} with a parametrized wind outflow profile with $S_\text{tot} = 80$ is also shown. 
The results in that case are close to our supersonic model. The main conclusion is that the yields of $^{92,94}$Mo and $^{96,98}$Ru in the model with a subsonic outflow are 1--2 orders of magnitude higher than with the supersonic wind, {for reasons explained below.} 

We also examine the amounts of $^{92}$Nb synthesized in these calculations.
Surprisingly, in the subsonic case it is produced in the necessary amounts to explain the solar system observations, while the standard lore says that this nuclide should not be made at all in the $\nu p$-process~\cite{Rauscher:2013}.
 We discuss this further in Secs.~\ref{subsec:physicsanalysis} and \ref{subsec:integrated}.

\subsection{Stages of the $\nu$p-process}\label{subsec:physicsanalysis}

\begin{figure*} 
    \centering
    \includegraphics[width = 0.49\textwidth]{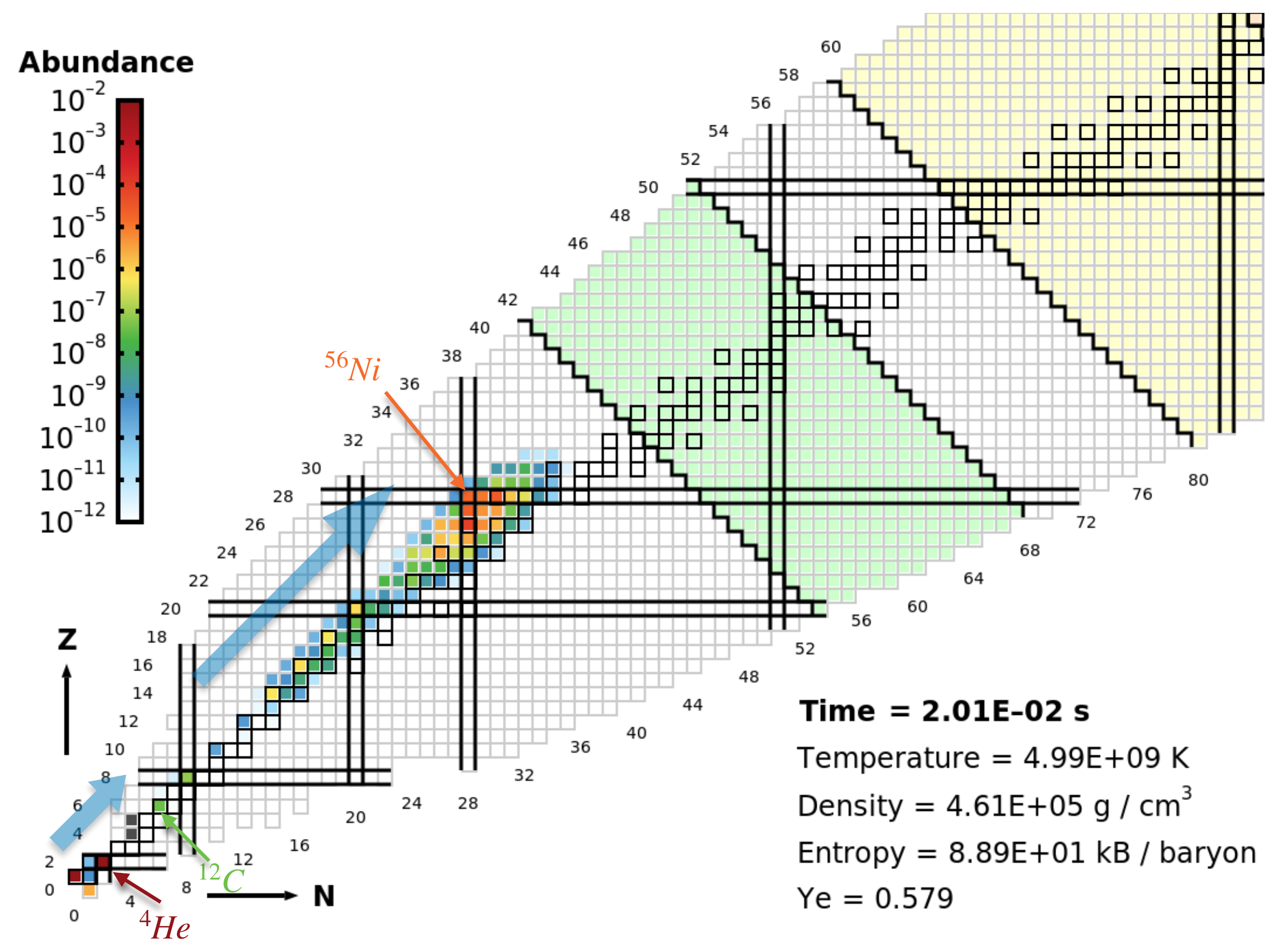}
    \includegraphics[width = 0.49\textwidth]{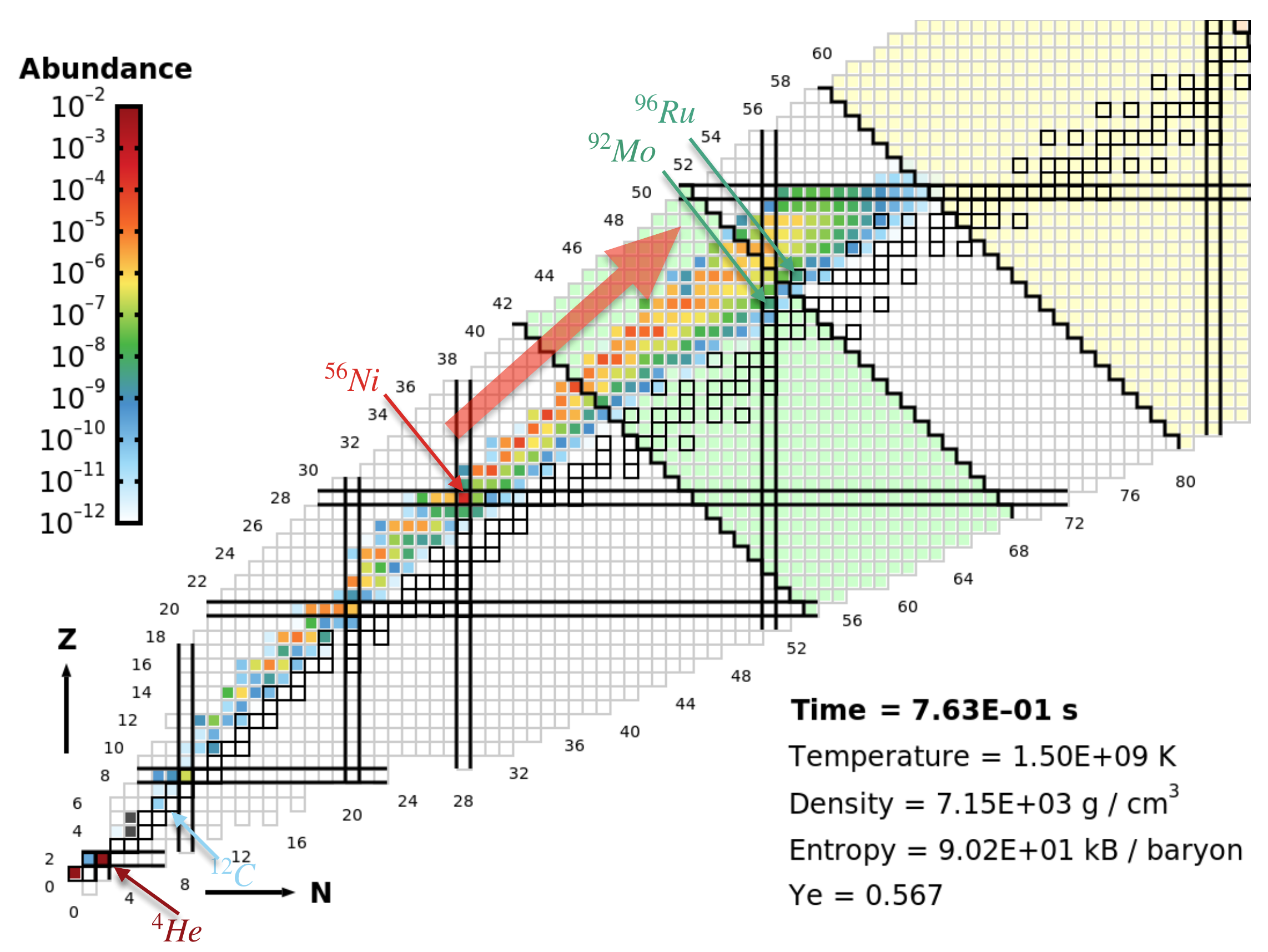}
    \includegraphics[width = 0.49\textwidth]{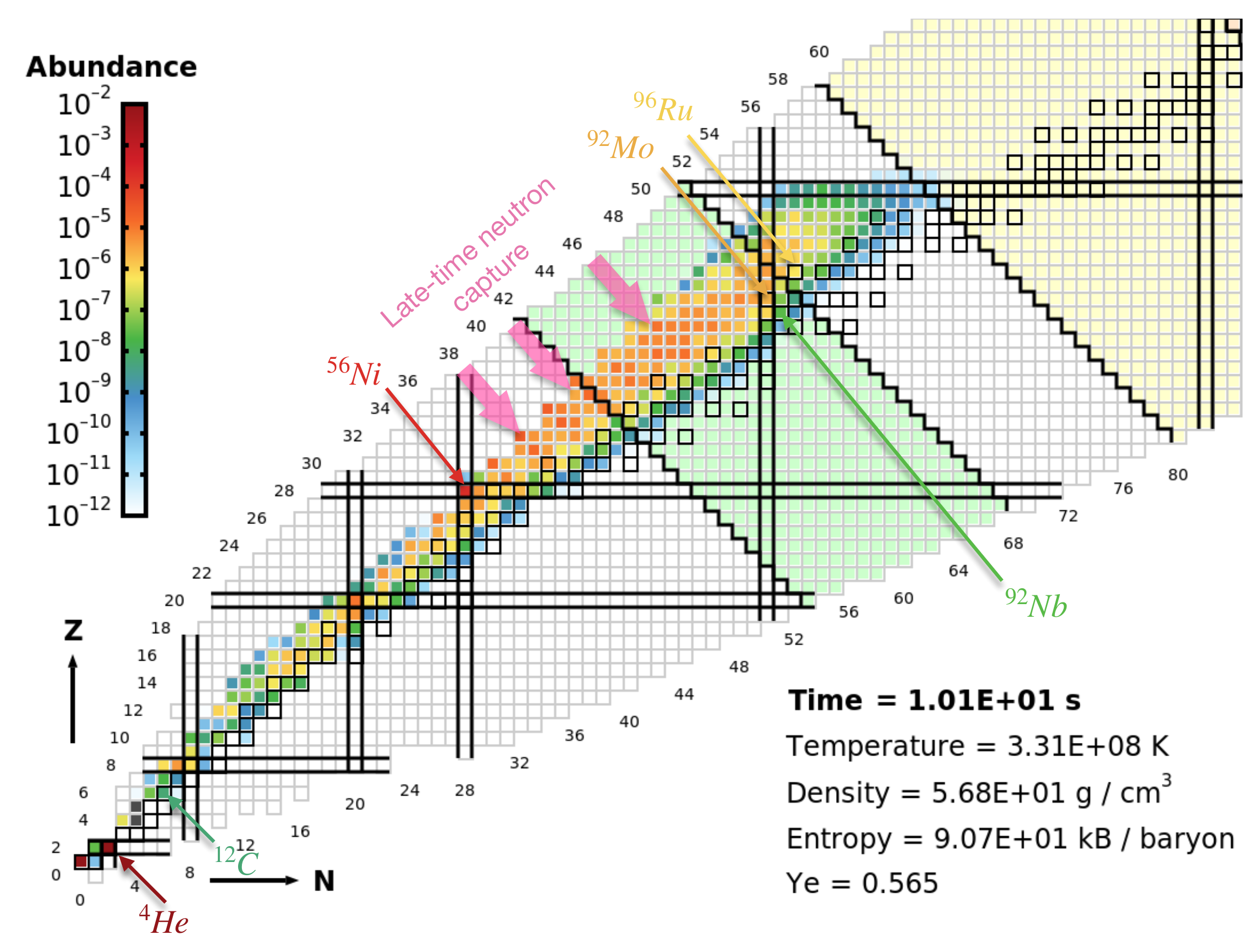}
    \includegraphics[width = 0.49\textwidth]{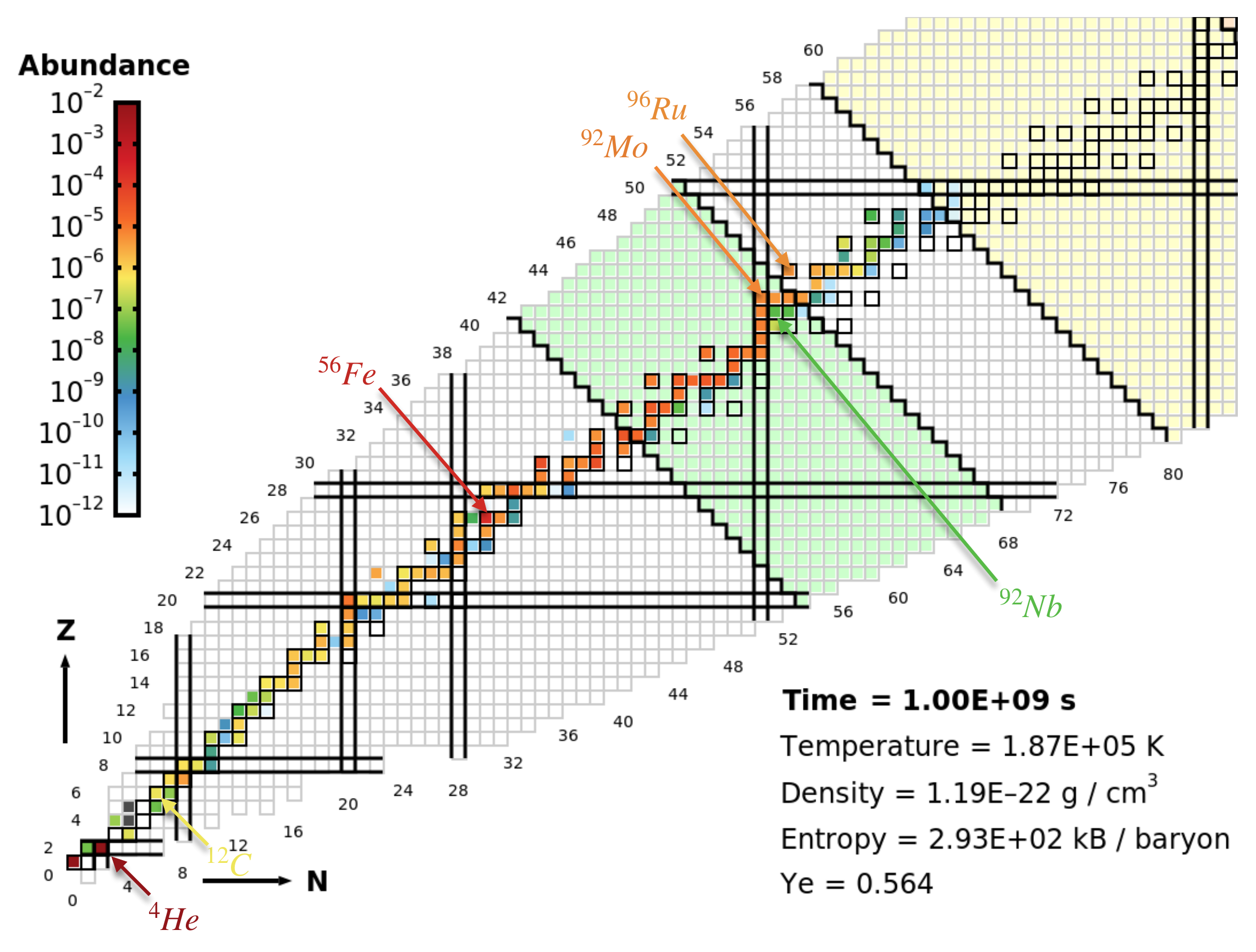}
    \caption{Stages of the $\nu p$-process in a proton-rich outflow ($13\,M_\odot$ progenitor model). \emph{Top left} (stage I): while $3\,\text{GK}\lesssim T \lesssim 6$\,GK, the outflow is made mostly of $\alpha$-particles and excess protons; the triple-$\alpha$ bottleneck ensures that only a small fraction of the $\alpha$s assemble into iron-group seeds.  \emph{Top right} (stage II): in the window $1.5\,\text{GK}\lesssim T\lesssim3$\,GK, proton capture, assisted by the capture of neutrons made in neutrino interactions, drives the formation of proton-rich isotopes in the mass range $64 \lesssim A \lesssim 105$. \emph{Bottom left} (stage III): below $T\sim1.5$ GK, proton capture becomes suppressed by the Coulomb barrier; neutrino interactions on free protons, however, continue neutron production; the resulting late-time $(n,\gamma)$ and $(n,p)$ reactions bring the composition closer to the valley of stability. \emph{Bottom right} (stage IV): by $t\sim10^9$ s, late-time $\beta$-decays complete the formation of (meta)stable isotopes. Notice the presence of shielded isotopes, including $^{92}$Nb, that could not be created by the $\beta$-decays only.}
    \label{fig:nupstages}
\end{figure*}

To gain insight into these numerical results, we analyze the physics of the $\nu p$-process. Fig.~\ref{fig:nupstages} shows the  snapshots in the neutron-proton plane from our simulation of the 13\,$M_\odot$ progenitor model.  The four stages are described in the caption.
%
%
Several conditions must be matched between these stages.
Let us investigate these in turn.

First, consider stages I and II. Efficiency of neutron production during stage II controls how high in atomic number the nucleosynthesis chain progresses. 
To produce sufficient amounts of $^{92,94}$Mo and $^{96,98}$Ru requires about ten neutrons per iron-group seed~\cite{Wanajo:2010mc}. 
Supersonic outflows tend to 
{underproduce neutrons compared to seeds}. In subsonic outflows, however, stage II occurs \textit{closer} to the PNS surface, where the material is subjected to a greater flux of neutrinos. This can be clearly seen, for example, in Fig.~\ref{fig:outflowprofiles} in Appendix~\ref{sec:nuspec}. As a result, up to three times as many neutrons are created as in the supersonic case. {The importance of \textit{distance} from the PNS (as opposed to simply the time spent) in stage II on neutron production has not been appreciated before.}

The factor counteracting this is the in-medium enhancement of the triple-$\alpha$ reaction~\cite{Beard:2017jpg,Jin:2020}. In proton-rich conditions, only a small fraction of the outflow, about 3--5\% by mass, progresses to the iron group nuclei, while most nucleons remain in the form of $\alpha$-particles (about 80\% by mass). The bottleneck occurs~\cite{Wanajo:2010mc} at the formation of $^{12}$C: $\alpha + \alpha + \alpha \rightleftharpoons {^8}\text{Be}^\ast + \alpha \rightleftharpoons \, ^{12}\text{C}^\ast$ create an excited state of $^{12}$C$^\ast$---the Hoyle state---that in most cases dissociates back into alpha particles. The formation of the ground state of $^{12}$C occurs by radiative decay of the Hoyle state and, in dense plasma, also  by collisions~\cite{Beard:2017jpg} with protons and neutrons in the medium. 

We find, however, that the neutron production rate during the second stage is so large in the subsonic case, that even with the in-medium enhancement of the triple-$\alpha$ rate the neutron-to-seed ratio is still optimal, so long as radiation entropy per baryon is $S\gtrsim65$ in the outflow (which is also used in \cite{Jin:2020}). In our self-consistent model, $S$ is not an independent parameter, but is set by the physical properties of the PNS. Specifically, $S$ is related to the gravitational potential at the PNS surface~\cite{Qian:1996xt}, which, given the physical constraints on the PNS radius at this stage of the explosion, translates into a lower bound on the PNS mass, $M_\text{PNS}\gtrsim1.7 \, M_\odot$. This is one of the key findings of our study and we will elaborate on it below.

We now turn to neutron production during stage III. We checked that, in our subsonic outflows, these neutrons do not drive the composition beyond the valley of stability to the neutron-rich side (cf.~~\cite{Arcones:2011zj}). This finding is nontrivial as there are no more independent physical parameters left to be adjusted in our model. 
 %
Even more remarkably, the number of the late-time neutrons found in our calculation---3--6 per seed---turns out to be optimal for solving the $^{92}$Nb origin problem. The standard lore says that isotopes such as $^{92}$Nb or $^{98}$Tc, which are shielded from $\beta$-decays by stable isotopes, cannot be produced in the $\nu p$-process at all. This assertion, which originates from the studies of the classical $rp$-process~\cite{DAUPHAS2003C287} and is reinforced by an unfortunate severe bug in the \texttt{Reaclib} library (see Appendix~\ref{sect:app_skynet}), turns out to be erroneous when applied to the $\nu p$-process. 
Both $^{92}$Nb and $^{98}$Tc are produced by neutron capture during stage III.

\subsection{Integrated yields} \label{subsec:integrated}

For quantitative comparisons with the solar system abundances, one should use time-integrated, rather than instantaneous, yields. We compute yields for each progenitor across a sequence of snapshots of the neutrino-driven outflow, corresponding to the trajectories of particles launched from the PNS surface at various times $t_\text{ps}$ after the shock revival (hereafter \lq\lq post-shock time\rq\rq). These yields are integrated, weighted by the corresponding mass loss rates $\dot M(t)$, as detailed below. For a nuclide $(A,Z)$, we define the time-averaged abundance $\langle Y_{A,Z} \rangle$ as:
\begin{equation}
    \langle Y_{A,Z} \rangle = \frac{\int \, Y_{A,Z}(t_\text{ps}) \, \dot{M}(t_\text{ps}) \, dt_\text{ps}}{\int \, \dot{M}(t_\text{ps}) \, dt_\text{ps}},
\label{eq:Ya_avg}
\end{equation}
where $Y_{A,Z}(t_\text{ps})$ is the abundance at a {post-shock time $t_\text{ps}$}, and $\dot{M}(t_\text{ps})$ is the corresponding mass outflow rate, obtained by solving the outflow equations from Appendix~\ref{subsec:outflow_calc}. 
The time-averaged mass fraction of the nuclide $(A,Z)$ is then $\langle X_{A,Z} \rangle = A \, \langle Y_{A,Z} \rangle$, provided the abundances are normalized so that the sum $\sum_{A,Z} A\,Y_{A,Z}$ over all nuclides adds up to 1.


The results can then be compared with the observed nuclide abundances in the solar system~\cite{Lodders:2003}. The isotopic \emph{production factor}~\cite{Wanajo:2010mc}
is defined as $f_{A,Z} = \langle X_{A,Z} \rangle / X_{A,Z}^\odot$, where $X_{A,Z}^\odot$ is the mass fraction of that isotope in the solar system. 
We take the solar system mass fractions of different isotopes from \cite{Lodders:2003}, where detailed isotopic abundances are tabulated from meteoritic studies of Carbonaceous Chondrites. In Ref.~\cite{Lodders:2003}, all the measured solar system abundances $Y_{A,Z}$ are normalized so that silicon has a total abundance of $10^6$. While converting these into mass fractions, one must use the appropriate normalization factor, which may be obtained by comparing the abundance of hydrogen ($2.431 \times 10^{10}$) to its observed mass fraction ($0.7110$). 
 The production factors are a means of comparing the \emph{relative abundances} of different isotopes synthesized in an environment. Isotopes with $f_{A,Z}$ not less than a tenth of the largest production factor $f_\text{max}$ can be considered to be co-produced in significant quantities~\cite{Wanajo:2010mc, Bliss:2018djg}.

\begin{figure*}[!htb]
\centering
\includegraphics[width = 0.49\linewidth]{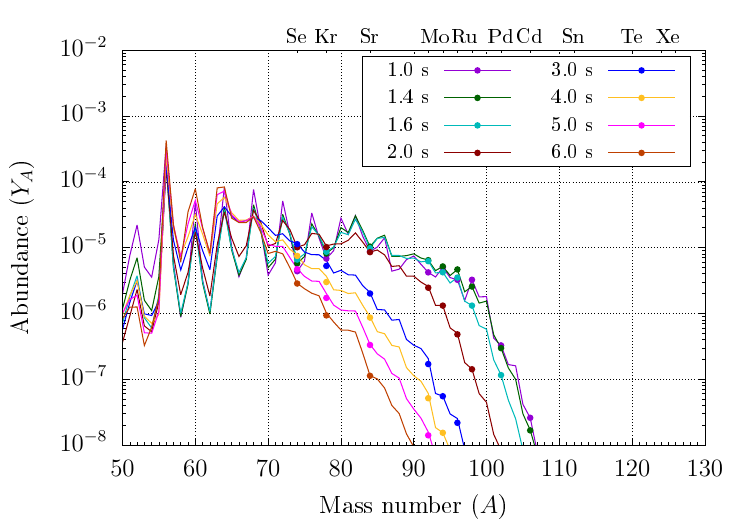}
\includegraphics[width = 0.49\linewidth]{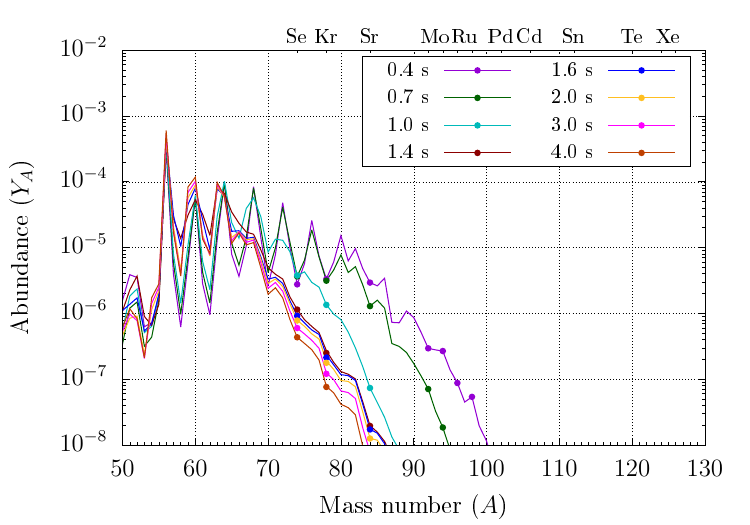} \\
\includegraphics[width = 0.49\linewidth]{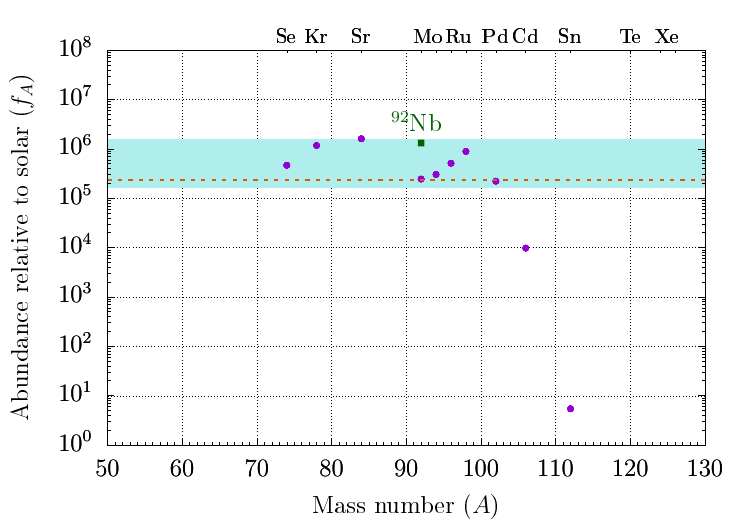}
\includegraphics[width = 0.49\linewidth]{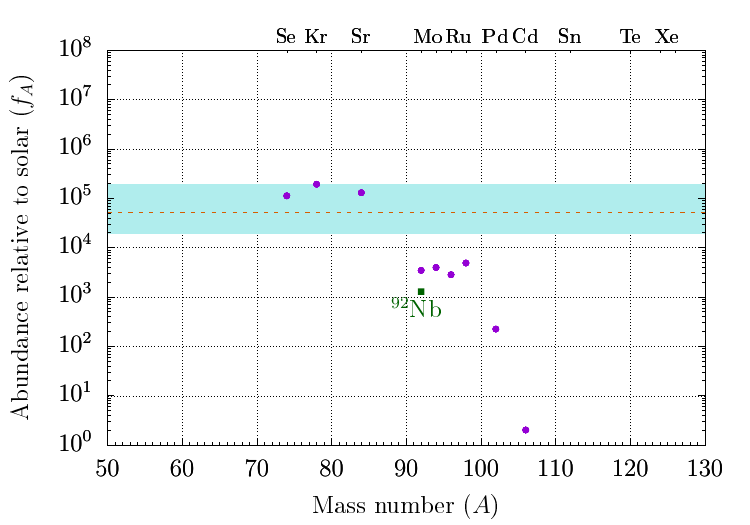} \\
\caption{\textit{Top panels}: Snapshots of nucleosynthetic yields from the explosions of $13$\,$M_\odot$ (left) and $9.5$\,$M_\odot$ (right) progenitors, labeled by the particle launch times. \textit{Bottom panels}: Integrated production factors (abundances relative to solar) of the $p$-nuclides of interest. 
The orange dashed line represents the minimum needed to explain the absolute abundances of these isotopes in the solar system~\cite{Woosley:1994ux,Wanajo:2010mc}. The colored band denotes the range of production factors between $f_\text{max}$ and $f_{\text{max}} / 10$, where $f_\text{max}$ is the highest production factor among all the $p$-nuclides. 
The subsonic outflow in the $13\,M_\odot$ model produces all the $p$-nuclides up to $^{102}$Pd within the colored band and above the  dashed line. 
It also produces the desired amount of $^{92}$Nb, whose integrated production factor here is normalized to $3\times10^{-3}$ of the solar $^{92}$Mo abundance, as suggested by studies of meteorites, cf.~\cite{Rauscher:2013,Lugaro:2016zuf,Iizuka:2016,Hibiya:2023}. 
The supersonic outflow in the $9.5M_\odot$ model significantly underproduces $p$-nuclides above $^{84}$Sr.}
\label{fig:integratedyields}
\end{figure*}

\begin{figure}
    \centering
    \includegraphics[width=0.80\textwidth]{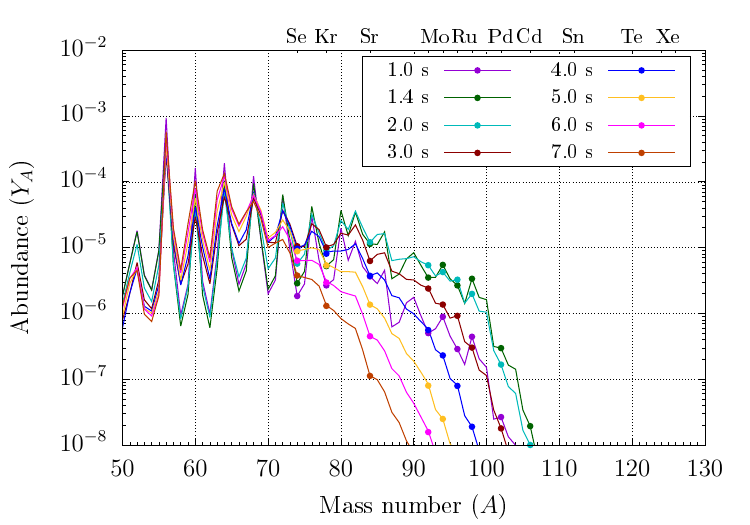}
    \includegraphics[width=0.80\textwidth]{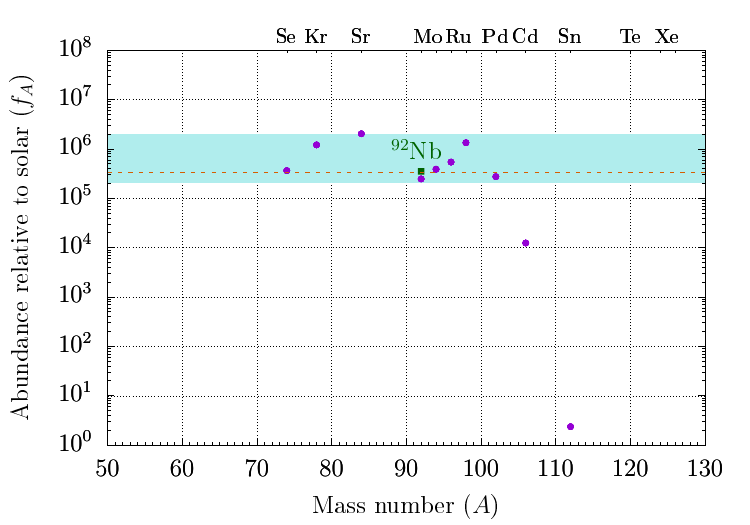}
    \caption{\textit{Top}: Time snapshots of nucleosynthetic yields for an $18$\,$M_\odot$ progenitor model. \textit{Bottom}: Integrated production factors (abundances relative to solar) of the $p$-nuclides of interest. The snapshot-by-snapshot yields as well as the integrated yields are qualitatively similar to those of the $13\,M_\odot$ progenitor calculation (Fig.~\ref{fig:integratedyields} left panels). Note that the {high yield time window} is reached at a \textit{slightly} later time for the $18\,M_\odot$ model (1.4--2\,s post-shock) as compared to the $13\,M_\odot$ model (1.0--1.6\,s), but still early enough for the mass-outflow rates, and hence the integrated yields, to remain sizeable.}
    \label{fig:18Mintegrated}
\end{figure}

The results of our integrated nucleosynthesis calculations are shown in Figs.~\ref{fig:integratedyields} and \ref{fig:18Mintegrated}. The left panels of Fig.~\ref{fig:integratedyields} depict our calculations for the $13\,M_\odot$ progenitor model, where the outflow remains subsonic for the duration of the explosion. The right panels show the corresponding results for the $9.5\,M_\odot$ progenitor model, where the outflow stays supersonic. Fig.~\ref{fig:18Mintegrated} shows the corresponding results for an $18\,M_\odot$ progenitor model. A qualitative agreement is observed between the $13$ and $18\,M_\odot$ models, demonstrating that the $\nu p$-mechanism works under subsonic outflow conditions for a broad range of progenitor masses.

The bottom panels in Figs.~\ref{fig:integratedyields} and \ref{fig:18Mintegrated} show the production factors of various $p$-nuclides of interest, for the different progenitor models.
The orange dashed line indicates the desired absolute amounts of each element~\cite{Woosley:1994ux, Wanajo:2010mc}, while the shaded band shows the range of production factors between $f_\text{max}$ and $f_{\text{max}} / 10$, where $f_\text{max}$ is the maximum production factor among the $p$-nuclides. Isotopes falling in this band are co-produced in sufficient quantities relative to the isotope with the highest production factor~\cite{Wanajo:2010mc, Bliss:2018djg}. We see that the subsonic outflow of the $13\,M_\odot$ progenitor model produces the correct amounts of all $p$-isotopes up to $^{102}$Pd. 
Also shown in each case are the production factors of $^{92}$Nb, normalized to 
$3\times 10^{-3}$ times the solar system meteoritic $^{92}$Mo abundance (since $^{92}$Nb itself, being unstable, is not observed in the present-day solar system). Analysis of meteoritic compositions, in combination with galactic chemical evolution models, have shown that the ratio of $^{92}$Nb/$^{92}$Mo at production has to be in the range $10^{-3}\text{--}10^{-2}$~\cite{Rauscher:2013, Lugaro:2016zuf, Iizuka:2016, Hibiya:2023}. 
This amount is sufficient to reconcile the solar system observations of $^{92}$Zr, into which this unstable nuclide eventually decays~\cite{Rauscher:2013,Lugaro:2016zuf,Iizuka:2016,Hibiya:2023}. We can also make predictions for other shielded isotopes that are produced by late-time neutrons during stage III of the $\nu p$-process, such as $^{98}$Tc, for which we predict $\sim 10^{-4}$ of the $^{98}$Ru abundance. This is consistent with the available upper limit on this nuclide in the solar system~\cite{DAUPHAS2003C287}, given its 6.1\,Myr half-life. {Note that $^{92}$Nb (and some $^{92}$Mo) could also be synthesized in mildly neutron-rich neutrino-driven outflows~\cite{Hoffman:1996}.}

For each progenitor, most of the $p$-nuclei with $A>80$ are produced in a narrow time window. This high-yield time window (HYTW) corresponds to the stage of the explosion when the material surrounding the hot bubble has temperature in the $1.5\,\text{GK}\lesssim T\lesssim3$\,GK range. The outflow upon deceleration reaches that temperature band and lingers in it for half a second or longer, allowing sufficient time for proton and neutron capture to drive $p$-isotope production~\cite{Wanajo:2010mc}. In addition to the time spent in this temperature range, the \textit{distance} to the PNS is another key factor. This distance is smaller in the subsonic case (e.g., Fig.~\ref{fig:outflowprofiles}), allowing for sufficient neutron production to make the $p$-isotopes up to $^{102}$Pd. 

The HYTW onset changes with progenitor mass---a heavier progenitor experiences it later than a lighter one, as the swept-up mass needs to be further diluted in the first case. However, the existence of the HYTW is always guaranteed. Thus, the $\nu p$ mechanism does not require progenitor mass fine-tuning---it is efficient in all sufficiently massive progenitors in which the outflows are subsonic. 
Another important point is that the HYTW occurs {within the first} 1--2\,s after the explosion is launched. At this time, the neutrino fluxes are still high and so is the amount of mass ejected. This makes it possible to generate sufficiently large \textit{absolute} $p$-nuclide yields.



To quantify these absolute isotopic yields, another metric often used for evaluating the efficacy of a nucleosynthetic process is the \emph{overproduction factor}~\cite{Woosley:1994ux,Wanajo:2010mc}.
It is defined as $f_{A,Z} \, \times ({M_\text{out}}/{M_\text{ejec}})$, i.e., the product of $f_{A,Z}$ and an astrophysical dilution factor which is the ratio of the total mass driven out in the neutrino-driven outflow ($M_\text{out}$) and the total mass ejected in the explosion ($M_\text{ejec}$). The outflow mass $M_\text{out} = \int\, \dot{M}(t_\text{ps}) \, dt_\text{ps}$ is obtained from the solutions of the outflow equations and the total ejecta mass is $M_\text{ejec} \approx M_\text{prog} - M_\text{PNS}$, where $M_\text{prog}$ is the progenitor mass. 
To explain the solar system abundance of an isotope, its overproduction factor in a supernova event must be $\gtrsim$ 10 \cite{Woosley:1994ux,Wanajo:2010mc}. 

One typically finds $M_\text{out} \sim 10^{-3}\,M_\odot$ and $M_\text{ejec} \sim 10\,M_\odot$ in a core-collapse supernova environment. For both the $13$ and $18 \,M_{\odot}$ models, we find the total outflow mass to be ${M_\text{out}} \approx 4.4\mbox{--}4.5 \times 10^{-4}\,M_\odot$. As seen in Figs.~\ref{fig:integratedyields} and \ref{fig:18Mintegrated}, for both these progenitors, $f_{A,Z} \, ({M_\text{out}}/{M_\text{ejec}}) \gtrsim 10$ is satisfied for $p$-nuclides up to $^{102}$Pd.
The values of the production factor $f_{A,Z}$ which correspond to an overproduction factor of 10 for each progenitor model are shown using an orange dashed line. In contrast, the integrated yields for the $A > 90$ nuclei in the $9.5\,M_\odot$ model (bottom right panel of Fig.~\ref{fig:integratedyields}) can clearly be seen to be deficient, both in an absolute sense and in comparison to the lighter $p$-nuclides.

\subsection{Protoneutron star properties} \label{sec:pnsmass}

\begin{figure}[bt]
\centering
\includegraphics[width = 0.8\linewidth]{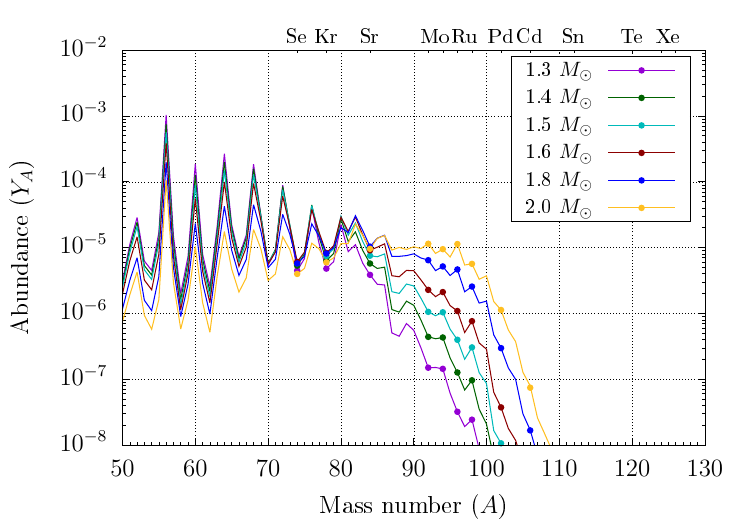}
\caption{Instantaneous $\nu p$-process yields for different protoneutron star masses, using a $13\,M_\odot$ progenitor model at 1.4\,s post shock launch. The radius of the PNS is kept fixed at 19 km. Clearly, the $\nu p$-process yields improve with higher masses. Explosions with {$M_\text{PNS} \gtrsim 1.7$\,$M_{\odot}$} are ideal sites.}
\label{fig:massdep}
\end{figure}

As already mentioned, the model with the successful $\nu p$ process has $M_\text{PNS} = 1.8\,M_\odot$. The physics underlying this choice is as follows. As the outflow expands, seed production is triggered when it reaches a temperature $\sim 6\,\text{GK}$, see Fig.~\ref{fig:evol}. The efficiency of seed production is controlled by the matter density at temperatures of $\sim 3 \text{--}6$ GK, since the triple-$\alpha$ reaction rates are strongly density dependent. The relationship between density and temperature is expressed by the entropy-per-baryon, $S$, which, in turn, is related to the gravitational potential near the PNS surface. 
The dependence of $S$ on the physical parameters---neutrino luminosity ($L_{\nu}$), average energy ($\langle E_\nu \rangle$), PNS mass ($M_\text{PNS}$) and radius ($R_\text{PNS}$)---is approximately $S \propto L_{\nu}^{-1/6} \, \langle E_\nu \rangle^{-1/3} \, R_\text{PNS}^{-2/3} \, M_\text{PNS}$ \cite{Qian:1996xt}. This scaling law shows that $S$ is mainly determined by the PNS mass and radius. Physically, the more massive and compact the PNS, the deeper the gravitational potential well, and the outflow is thus required to have more radiation energy per baryon, $T^4/n\propto TS$, to become unbound. 

%
Next, we need to consider the physical range of $R_{\rm PNS}$. Most studies of the nuclear equation of state focus on the radius of a cold neutron star, which is typically in the range 10--12\,km. However, most of the yields are produced in the HYTW, which occurs at $\mathcal O(1\mbox{--}2 \,\mathrm{s})$ after the shock revival. At such times, the radius is larger, because the PNS still has a significant amount of trapped thermal energy and lepton number. 
Modern simulations report values in the range $R_\text{PNS} \sim 17\mbox{--}22\,\text{km}$ at 1--2\,s post-explosion~\cite{Fischer:2009,Roberts:2016rsf,Burrows:2019rtd,Bollig:2020phc,Vartanyan:2018iah}. 
Given this, to obtain the required value of $S$, the PNS mass has to be in the range {$M_{\rm PNS}\gtrsim1.7 M_\odot$}. This leads to not only the required absolute amounts of $p$-nuclides up to $^{102}$Pd, but also the desired pattern of relative abundance ratios of $^{92,94}$Mo and $^{96,98}$Ru. Instantaneous $p$-isotope yields for different PNS masses are shown in Fig.~\ref{fig:massdep}. 

Our findings align nontrivially with modern simulations. The PNS mass is controlled by the amount of matter accreted onto the PNS post collapse.  For light progenitors with masses of $9\mbox{--}10\,M_\odot$ the explosion is achieved quickly and the resulting baryonic PNS mass is close to the Chandrasekhar value~\cite{Burrows:2019rtd}. For more massive progenitors, however, modern multi-dimensional simulations find that extended accretion is not only possible, but a common feature. For progenitors of $13\mbox{--}20\, M_\odot$, one finds PNS {masses} in the range $1.6\mbox{--}2.0 \, M_\odot$~\cite{Bollig:2020phc, Burrows:2020qrp}. This is also supported by the measurements of pulsars in binary- and multiple-star systems that the upper limit on the neutron star mass is at least $\sim 2\,M_\odot$~\cite{ozel:2012,Ozel:2016oaf,Antoniadis:2016hxz}. Nearly 20\% of the measured NS masses are $1.8\,M_\odot$ or greater, and it has been suggested that the mass distribution is bimodal, with a secondary peak centered at $M_\text{NS} \sim 1.8\,M_\odot$~\cite{Antoniadis:2016hxz}. Note that the masses of cold neutron stars are $5\text{--}10$\% lower than during the HYTW when a significant amount of trapped thermal energy and lepton number still remains in the PNS.



\begin{figure}[htb]
\centering
\includegraphics[width = 0.8\linewidth]{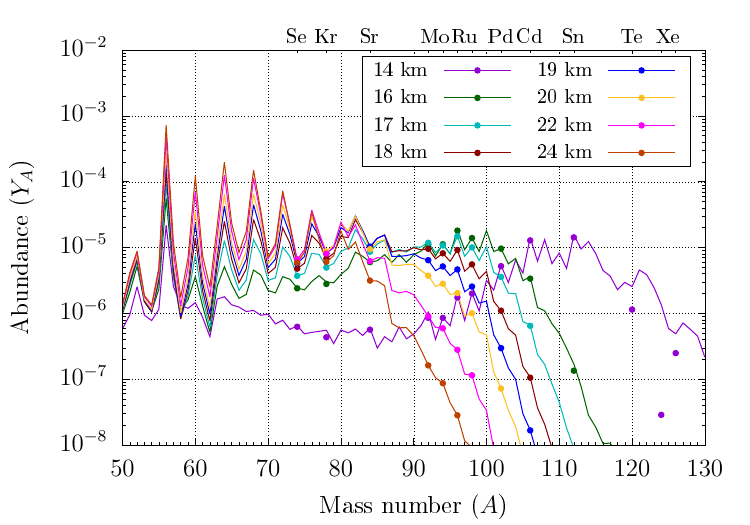}
\caption{Instantaneous $\nu p$-process yields for different protoneutron star radii ranging from 14--24 km, using a $13\,M_\odot$ progenitor model at 1.4\,s post shock launch. The mass of the PNS is kept fixed at $1.8\,M_{\odot}$. Clearly, more compact objects are better suited for more efficient $\nu p$-process.}
\label{fig:radialdep}
\end{figure}

To complete this argument, in Fig.~\ref{fig:radialdep} we report the variation of $\nu p$-process yields with PNS radius, with the PNS mass kept fixed at $1.8 \, M_{\odot}$. Compact objects with $R_\text{PNS} \sim 14$ km can produce $p$-nuclides even up to $A \sim 120\mbox{--}130$, while for larger radii $R_\text{PNS} > 20$ km, the yields are negligible beyond $^{84}$Sr. 
The observed solar abundances are well reproduced for $R_\text{PNS}\sim18\text{--}20$ km.


\section{Concluding remarks}

We have used a suite of one-dimensional semi-analytical models to survey the parameter space of the $\nu p$-process in CCSN. We uncovered a regime in which the conditions reproduce the solar system observations of $p$-nuclei up to $^{102}$Pd, including the famous $p$-isotopes of Mo and Ru, and avoid all the objections raised in the literature. Several pieces fall into place in a nontrivial way. First, the desired abundance pattern is achieved for all parameters in their physically plausible ranges. Second, for sufficiently heavy progenitors, modern simulations find that the neutrino driven outflows are subsonic and the PNS is heavy due to an extended accretion stage. Third, late-time neutrons not only do not drive the composition to the neutron-rich side, but also produce the shielded $^{92}$Nb in the correct abundance.  
All these results motivate and guide future investigations. It is desirable to assess the impact of different nuclear equations of state and of neutrino flavor transformations, including collective oscillations that may occur above the PNS surface. Above all, the regime found here deserves further investigation with modern multi-dimensional simulations.  

\begin{acknowledgments}
The majority of this work was carried out at SLAC, where it was supported by the U.S. Department of Energy under contract number DE-AC02-76SF00515. We also thank the Kavli Institute for Theoretical Physics (KITP), the Institute for Nuclear Theory (INT) at the University of Washington, and {the Mainz Institute for Theoretical Physics (MITP)}, where parts of this work were completed and preliminary results  presented. This research was supported at KITP by the National Science Foundation under Grants No. NSF PHY-1748958 and PHY-2309135, at INT by the U.S. Department of Energy grant No. DE-FG02-00ER41132, {and at MITP by the Cluster of Excellence PRISMA+ (Project ID 39083149)}. {AVP also acknowledges support from the U.S. Department of Energy under contract number DE-FG02-87ER40328 at the University of Minnesota, as well as partial travel support from the Network for Neutrinos, Nuclear Astrophysics, and Symmetries (N3AS), funded by the National Science Foundation under grant no. PHY-2020275.} {The work of PM is supported in part by the Neutrino Theory Network Program Grant under award number DE-AC02-07CHI11359.} We gladly acknowledge C.~Horowitz, G.~Fuller, S.~Reddy, R.~Surman, and G.~Martinez-Pinedo for useful discussions and I.~Padilla-Gay for helpful feedback on the manuscript.
\end{acknowledgments}

\appendix

\section{Outflow calculations} \label{subsec:outflow_calc}

Each iteration of the nucleosynthesis calculation requires modeling matter trajectories through the stages depicted in Fig.~\ref{fig:nupstages}. The matter starts out close to the PNS surface where nuclear statistical equilibrium prevails and is followed to a large radius until neutron capture ceases. This covers stages I through III in Fig.~\ref{fig:nupstages}; the last stage involves only beta-decays, which does not require detailed trajectory modeling. The duration of stages I through III is comparable to the timescales on which the neutrino luminosities decay and the matter profile changes in the developing explosion. Therefore, the nucleosynthesis trajectories are not instantaneous snapshots of the matter profile. Yet, the physics of the problem affords an important simplification: the outflow can be divided into two segments. The first segment describes the acceleration of the material by neutrino heating near the PNS surface and its deceleration by the surrounding slowly expanding material. The second one involves the subsequent evolution upon joining the homologous expansion of the material behind the front shock, following $v \sim \text{const}$ (cf.~\cite{Fischer:2009,Wanajo:2010mc}). The first segment is quick enough ($\lesssim 1$ s) to treat the outer boundary condition and the neutrino fluxes as approximately constant in time. The second one has time-varying $\rho$ and $T$ dictated by the expansion of the front shock, as well as neutrino luminosities variable on the PNS cooling timescale. This framework allows us to explore the yields as a function of physical conditions while avoiding using full SN simulations.

The first segment has been modeled numerically, by solving a set of steady-state, spherically symmetric equations. Our approach is based on the classical treatment of neutrino-driven outflows~\cite{Duncan1986,Qian:1996xt}, but with three important differences. First, we take into account general relativistic corrections~\cite{Cardall:1997bi,Otsuki:1999,Thompson:2001}. Second, we incorporate  the change in the effective number of relativistic degrees of freedom $g_\star$, which occurs as the outflow cools through the temperature of $e^+e^-$ annihilation between the PNS surface and the outer edges of the hot bubble~\cite{Friedland:2020ecy}. Last, we consistently treat the far boundary condition, using the methodology described in~\cite{Friedland:2020ecy}, which treats subsonic and supersonic regimes, and the transition between the two, in a unified fashion. With this, and using
the Schwarzschild background, we have~\cite{Cardall:1997bi}

\begin{align}
   \left( \frac{v}{1 - v^2} \right)\left(1  -  \frac{v_s^2}{v^2} \right)\frac{dv}{dr} \quad &= \quad
        \frac{2 v_s^2}{r}  - \frac{G M}{r^2} \frac{(1 - v_s^2)}{\left(1- \frac{2 G M}{r}\right)} 
        - \beta\frac{\dot{q}}{v y (1 + 3 v_s^2)}
        \label{eq:fluidacceleration2_full}, \\
  \frac{\dot{q}}{1 + 3 v_s^2} \quad &= \quad 
        v y \frac{d}{dr} \left[\ln{(1 + 3 v_s^2)} - \frac{1}{2} \ln{(1 - v^2)} 
        + \frac{1}{2}\ln{\left( 1 - \frac{2 G M}{r}\right)}\right]
        \label{eq:dIdr}, \\
      v y\frac{dS}{dr} \quad &= \quad \frac{\dot{q}m_N}{T}. 
      \label{eq:qdotdS} 
\end{align}
Here $v$ is the coordinate velocity of the fluid element, $v_s$ is the sound speed, $S$ is radiation entropy per baryon and $y = \sqrt{(1 - 2GM/r)/(1 - v^2)}$. Natural units are assumed, so that the speed of light $c=1$. Furthermore, 
\begin{eqnarray}
v_s^2 &=& \frac{S T}{4 m_N}\left(1+ \left( 3+\frac{d \ln g_\star}{d \ln T}\right)^{-1}\right),\\
\beta &=& \frac{1}{4}\left(1+ \left( 3+\frac{d \ln g_\star}{d \ln T}\right)^{-1}\right).
\end{eqnarray} 

The quantity $\dot{q}$ is the specific energy deposition rate, which has three main contributions: heating due to $\nu_e$ and $\bar\nu_e$  absorption on nucleons  ($\dot{q}_{\nu N}$), cooling due to neutrino emission in $e^+$ or $e^-$ absorption on nucleons ($\dot{q}_{eN}$), and cooling due to $e^+e^-$ pair annihilation into neutrino-antineutrino pairs ($\dot{q}_{e^+e^-}$). Numerically, these rates are~\cite{Qian:1996xt,Cardall:1997bi}:
\begin{align}
  \dot{q}_{\nu N} \quad &= \quad
            9.84 N_A \left[(1 - Y_e)L_{\nu_e,51} \varepsilon_{\nu_{e,\mathrm{MeV}}}^2 + Y_e L_{\bar{\nu}_e,51} \varepsilon_{\bar{\nu}_{e,\mathrm{MeV}}}^2 \right]
            \times \frac{1 - g_1(r)}{R_{\nu,6}^2} \Phi(r)^6 ~ \mathrm{MeV ~s^{-1} g^{-1}},
        \label{eq:nuN} \\[\jot]
 \dot{q}_{e N} \quad &= \quad 2.27 N_A T_{\mathrm{MeV}}^6 ~ \mathrm{MeV ~s^{-1} g^{-1}},
 \label{eq:eN} \\[\jot]
      \dot{q}_{e^+e^-} \quad &= \quad 0.144 N_A \frac{T_{\mathrm{MeV}}^9} {\rho_8} ~ \mathrm{MeV ~s^{-1} g^{-1}}.
      \label{eq:epluseminus} 
\end{align}
where $N_A$ is the Avogadro number, $R_{\nu,6}$ is the neutrinosphere radius in units of 10$^6$ cm, $L_{\nu,51}$ is the neutrino luminosity in units of 10$^{51}$ erg/s, $\varepsilon = \sqrt{\langle E^3 \rangle/ \langle E \rangle}$ and  $\epsilon = \langle E^2 \rangle /\langle E \rangle$. The redshift factor $\Phi(r)$ is given by
\begin{equation} \label{eq:redshift}
    \Phi(r) = \sqrt{\frac{1 - 2G M / R_{\nu}}{1 - 2 G M / r}}
\end{equation}
and the factor ($1 - g_1(r)$) represents the effect of the gravitational bending of the neutrino trajectories, 
\begin{equation}
    g_1(r) = \sqrt{1 - \left(\frac{R_{\nu}}{r}\right)^2 \left( \frac{1 - 2G M / r}{1 - 2 G M / R_{\nu}}\right) }.
\end{equation}


To completely specify the problem, one should  supply (i) the luminosites and energy spectra of the electron neutrinos and antineutrinos, (ii) the properties of the protoneutron star---its mass and radius, and (iii) the appropriate physical boundary conditions at the starting and ending radii. The three boundary conditions we impose are the temperature and density at the PNS surface and, crucially, the pressure of the surrounding material, $P_\text{far}$. This surrounding pressure is considered fixed for a given snapshot, as discussed earlier, but is varied between the snapshots. It is dictated by the density of the surrounding material, $\rho_s$ and its radiation entropy per baryon, $S_{s}$. The density, in turn, is controlled by the amount of matter plowed up by the expanding front shock, $M_\text{plow}(t)$ and the radius of the front shock, $R_s(t)$:
\begin{equation}
    \rho_{s}(t) \sim M_\text{plow}/[(4 \pi/3)R_s^3], \qquad R_s = v_s t.
    \label{eq:density_evol_postshock}
\end{equation}
The value of $M_\text{plow}(t)$ depends on the progenitor density profile and the \emph{mass cut} (the position of the mass element that separates ejected and accreted material). In our calculations, we take progenitor profiles from \cite{Sukhbold:2015}. 

In order to convert $\rho_s(t)$ to $P_\text{far}(t)$, we impose pressure continuity across the boundary of the hot bubble. Let $\rho_{f}$ be the far density of the outflow, $S_f$ be the radiation entropy in it, and $S_{s} \sim 6$ be the radiation entropy of the surrounding material. The equality of the radiation pressures then implies that temperature is continuous, while the density jumps by the ratio of the radiation entropies: $\rho_{s}/\rho_f = S_f/S_{s} \sim 12$. This density jump at the edge of the high-entropy hot bubble region is seen in 1D simulations and is commonly known as a contact discontinuity.

{The system is near-critical, in a sense that both supersonic and subsonic outflow regimes are possible, depending on the interplay of the physical parameters, namely, the properties of the PNS, the neutrino fluxes, and the progenitor density profile. To avoid any confusion, we stress that we do not do any fine-tuning of the outflow parameters to the transition between subsonic and supersonic regimes (e.g., making the shock position be close to the sonic point, see~\cite{Wanajo:2010mc}).
}

The system is solved numerically following the procedure developed in~\cite{Friedland:2020ecy}. The boundary condition on the surrounding pressure is matched by ``shooting'' the initial velocity in the subsonic case and the position of the shock in the supersonic case. This procedure continuously covers the entire range of physical possibilities and self-consistently determines the nature of the outflow.


The second segment of the trajectory, governed by the front-shock expansion, must be modeled for several seconds, during which neutron production is still taking place. During this time, the distribution of the lighter $p$-elements evolves (the peaks and troughs are smoothed by neutron capture), as well as the synthesis of $^{92}$Nb takes place. The key physics observation is that the material behind the front shock is expanding approximately homologously, so that the velocities as a function of $r$ follow the Hubble law. This is so because the entire region is in causal contact and the effects of gravity for $r>10^3$ km on the density distribution are subdominant. The outflow material joins this expansion when the velocity computed in the first segment falls to the value predicted by the Hubble law. Typically, the transition happens at around $r\sim10^3$ km, where the velocity is a fraction of that of the front shock. For example, when the front shock is at $R_s\sim 10^4$ and has $v_s\sim10^4$ km/s, the transition velocity at $10^3$ km is $10^3$ km/s. The speed of the matter element remains a fixed fraction of the front shock speed, and its density is governed by Eq.~(\ref{eq:density_evol_postshock}). This physical model is in good agreement with the results in \cite{Fischer:2009}, as well as with the empirical parametrization of simulation results employed in \cite{Wanajo:2010mc}. {Detailed calculations regarding the model and the formation of subsonic or supersonic outflows can be found in \cite{Mukhopadhyay:2022yrd, Friedland:2020ecy}.}


The results of this procedure are shown in Figs.~\ref{fig:integratedyields} and \ref{fig:18Mintegrated}, where we {show the nucleosynthesis calculations along different time snapshots}. The heating rates, $\dot{q}_i$ have been computed using an exponentially decreasing luminosity: $L_{\nu_\gamma}(t) = L_{\nu_\gamma}(t_0) \, e^{-(t-t_0)/\tau}$, with a time constant $\tau = 3\,$s, as described below.
Our model passes a number of qualitative and quantitative validation tests, when compared to the full 1D explosion simulations in the literature~\cite{Fischer:2009, Arcones:2006}. At the basic level, heavier progenitor stars give rise to higher swept-up masses, resulting in higher surrounding pressures and subsonic outflows. 
In detail, our model reproduces intricate features of transient shocks, where termination shocks appear during a small time window, and then disappear, in good agreement with the $\sim$ 10 M$_{\odot}$ progenitor simulations of \cite{Fischer:2009}. 


\section{Neutrino luminosities and spectra} \label{sec:nuspec}

Neutrinos streaming from the cooling PNS determine both the heating rate $\dot q$ in the outflow equations and the neutron production rate for the nucleosynthesis network. In this work, we adopt a ``pinched Fermi-Dirac'' parametrization for the neutrino distributions~\cite{Keil:2002in}:
\begin{equation}
    f_{\nu_\gamma} \propto \frac{1}{e^{E/T_{\nu_\gamma} - \eta_{\nu_\gamma}} + 1},
\end{equation}
where $\nu_\gamma \in \{\nu_e, \bar\nu_e\}$. The distribution for each species is specified with three parameters: the effective neutrino temperature $T_{\nu_\gamma}$ and the pinching parameter $\eta_{\nu_\gamma}$---which together set the spectral shape---and the luminosity $L_{\nu_\gamma}$, which fixes the flux normalization. These parameters are independent and not related by the black-body formula, reflecting the complex dynamics of neutrino decoupling.

We choose an exponentially decreasing luminosity, as described in Appendix~\ref{subsec:outflow_calc}. It is important to account for this decrease in luminosity within each nucleosynthesis snapshot calculation, in order to 
correctly estimate the number of neutron captures during the third stage in Fig.~\ref{fig:nupstages} ($T < 1.5$\,GK). 
An overestimation of late-time neutron captures could potentially turn material that is proton-rich at 1.5\,GK into a neutron-rich state afterward~\cite{Arcones:2011zj}. This effect would be expected to be stronger in subsonic outflows because of the greater relative proximity to the PNS, or at high entropies because of the relative under-abundance of seeds relative to the number of late-time neutrons (e.g., the $S = 140$ case in Ref.~\cite{Jin:2020}).

Most of the yields are generated in the first $1\mbox{--}2$\,s after shock launch, in the HYTW (see Sec.~\ref{subsec:integrated} for details). The spectral parameters $T_{\nu_\gamma}$ and $\eta_{\nu_\gamma}$ are generally found in simulations not to vary significantly in this window (e.g.,~\cite{Huedepohl:2009wh}). Therefore, for simplicity, we hold these parameters, and the protoneutron star radius $R_\text{PNS}$, fixed throughout our calculation. The final outcome is thus not particularly sensitive to variations of these physical parameters beyond $\sim 2$\,s post-shock.

The neutrino and PNS parameters used in our calculations are presented in Table~\ref{table:nuparams}. The ratios of the neutrino and antineutrino luminosities ensure $Y_e \simeq 0.6$ in the outflow, leading to an optimal $\nu p$-process (cf.~\cite{Wanajo:2010mc}). For the $9.5\,M_\odot$ model, we use luminosities that are $30\%$ higher than for the heavy progenitor calculations, as well as a smaller pinching parameter for $\bar\nu_e$. These choices enabled us to obtain a supersonic outflow solution with a wind-termination shock that qualitatively resembled the parametrized profiles in~\cite{Jin:2020}, facilitating comparison of results.  
Neutrino spectra are known to vary sensitively between simulations, and can be further impacted by other physics such as flavor oscillations. While the values chosen here (and in other literature, e.g., \cite{Wanajo:2010mc,Jin:2020}) are physically plausible, a follow-up investigation of the impact of different spectra is highly motivated. 

\setlength{\tabcolsep}{5pt}
\begin{table}[ht]
\centering
\renewcommand{\arraystretch}{1.2}
\begin{tabular}{| l | r | r |}
\hline
\textit{Parameter}                  & \textit{Value}                    & \textit{Value} \\ 
                                    & $13\,M_\odot$ simulation      & $9.5\,M_\odot$ simulation \\ 
\hline \hline
 $ L_{\nu_e} (1\,\mbox{s}) $        &   $7 \times 10^{51}$\ erg/s       & $9.1 \times 10^{51}$\ erg/s \\ 
 $ L_{\bar\nu_e} (1\,\mbox{s}) $    &   $5.74 \times 10^{51}$\ erg/s    & $7.46 \times 10^{51}$\ erg/s \\ 
 $ \langle E_{\nu_e} \rangle $      &   $9.7$\ MeV                      & $9.7$\ MeV  \\ 
 $ \langle E_{\bar\nu_e} \rangle $  &   $11.7$\ MeV                     & $11.7$\ MeV \\ 
 $ \eta_{\nu_e} $                   &   $2.1$                           & $2.1$ \\ 
 $ \eta_{\bar\nu_e} $               &   $1.5$                           & $0.7$ \\
 \hline
$ M_\text{PNS} $                    &   $1.8$\ $M_\odot$                & $1.4$\ $M_\odot$\\ 
$ R_\text{PNS} $                    &   $19$\ km                        & $19$\ km \\
$Y_{e,0}$                           &   0.6                             & 0.6 \\ 
\hline \hline
\end{tabular}
\caption{Neutrino and protoneutron star parameters used in the $13\,M_\odot$ and $9.5\,M_\odot$ simulations presented in Sec.~\ref{sec:results}. 
The luminosities at time $t$ were taken to be $L_{\nu_\gamma}(t) = L_{\nu_\gamma}(t_0) e^{-(t-t_0)/\tau}$, with $\tau = 3$\,s, whereas the average energies $\langle E_\nu \rangle$ and pinching parameters $\eta_\nu$ were held fixed. 
}
\label{table:nuparams}
\end{table}

Note that the neutrino spectral parameters values specified in Table~\ref{table:nuparams} are from the point of view of a distant observer. In the outflow and nucleosynthesis calculations, the neutrino fluxes and energies in the vicinity of the PNS are calculated relative to their values at infinity using the appropriate gravitational blue-shift factors (powers of $(1 - 2GM/r)^{-1/2}$). The geometric dilution of the neutrino fluxes is assumed in accordance with an isotropic emission (in outward directions) from a spherical surface with radius $R_{\nu}$, and is given by
\begin{equation} \label{eq:geomdil}
    \mathcal{D}(r) = \frac12 \left( 1 - \sqrt{1 - \frac{R_{\nu}^2}{r^2}} \right).
\end{equation}

\begin{figure*}[!htb]
\centering
\includegraphics[width = 0.49\linewidth]{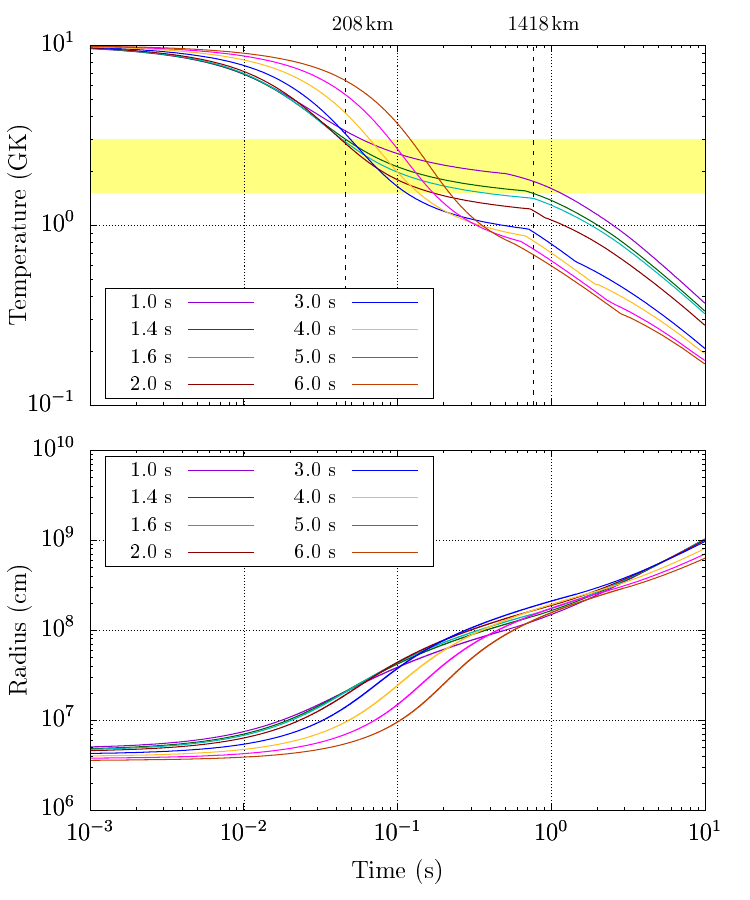}
\includegraphics[width = 0.49\linewidth]{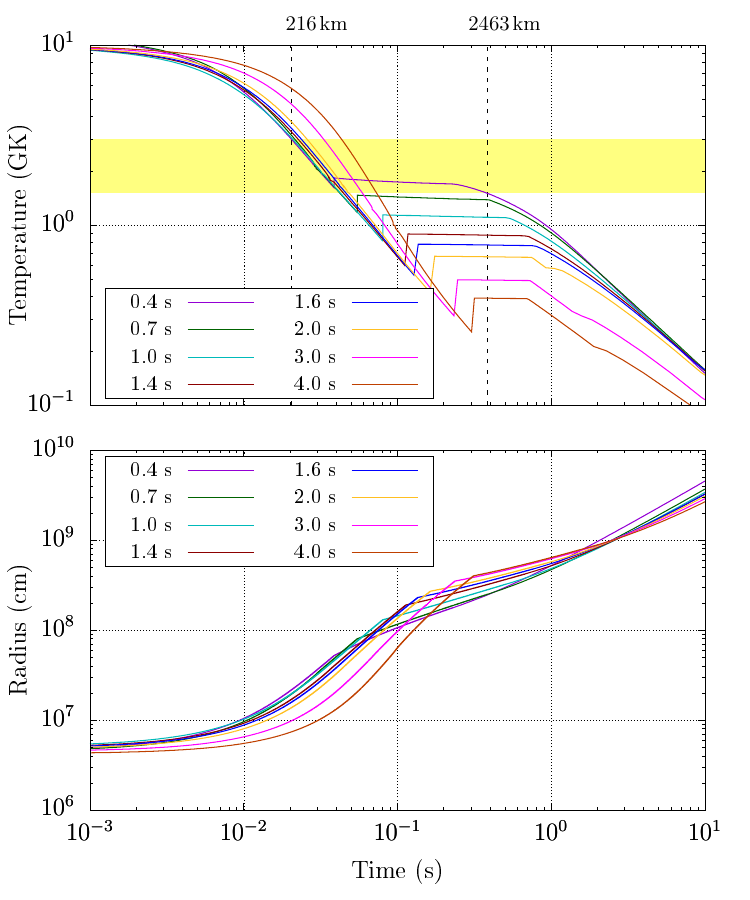} \\
\caption{Outflow profiles from the explosions of $13$\,$M_\odot$ (left) and $9.5$\,$M_\odot$ (right) progenitors, labeled by the particle launch times. \textit{Top panels}: Outflow temperature vs time post-launch, for outflows launched at different times. Also shown in the top panels in each case are (i) the temperature band $3\,\text{GK} > T > 1.5\,\text{GK}$ (yellow), during which $(p,\gamma)$, $(n,p)$, and $(n,\gamma)$ reactions together conspire to synthesize $p$-rich nuclei in the outflow; and (ii) the radial distances from the PNS (dashed vertical lines) where specific representative snapshots---1.4\,s for the $13\,M_\odot$ model and 0.4\,s for the $9.5\,M_\odot$ model---enter and exit this temperature band. \textit{Bottom panels}: Radial distance vs time post-launch, for the outflows launched at different times.}
\label{fig:outflowprofiles}
\end{figure*}

In the outflow calculations, we take $R_\nu = R_\text{PNS}$. The outflow profiles computed using the framework outlined in Appendix~\ref{subsec:outflow_calc}, with neutrino luminosities and spectra as specified above, are shown in Fig.~\ref{fig:outflowprofiles} for the $13\,M_\odot$ (left) and $9.5\,M_\odot$ (right) progenitor models, across a sequence of launch times.

\section{Nucleosynthesis calculations with SkyNet}
\label{sect:app_skynet}

The nucleosynthesis network calculations were carried out using \texttt{SkyNet}~\cite{Lippuner:2017tyn,skynet}, an open source reaction network, which adopts the majority of nuclear reaction rates from the \texttt{Reaclib} library, version 2.2~\cite{reaclib2010,reaclibweb}. Starting with the \texttt{special\_reaction\_library\_rebased} branch of \texttt{SkyNet}, two principal modifications were made: (i) the triple-$\alpha$ reaction rate was boosted by in-medium enhancement effects~\cite{Beard:2017jpg, Jin:2020}; and (ii) a crucial bug in \texttt{Reaclib} describing the decay of $^{92}\text{Nb}$ was corrected. 

The modified triple-$\alpha$ rate was implemented using the publicly available code~\cite{triplealpha} from the authors of Ref.~\cite{Jin:2020}, with a number of further modifications, including the general-relativistic (GR) corrections and time-dependent neutrino luminosities, both as described above. The GR corrections are found to increase the entropy per baryon by about $\Delta S \sim 5$, which decreases seed production during stage I, as desired. 

The \texttt{Reaclib} issue is easy to state: any $^{92}$Nb made in \texttt{SkyNet} wasfound to undergo a $\beta$-decay $^{92}\text{Nb} \, \rightarrow \, ^{92}\text{Mo}$  on a timescale of $\mathcal{O}(100\,\text{s})$. This behavior is incorrect: this decay channel is strongly forbidden and is neither observed nor expected~\cite{Baglin:2012aio}. The bug drives any $^{92}$Nb produced back to $^{92}$Mo, thus unfortunately reinforcing the shielding argument~\cite{Rauscher:2013}. Instead, $^{92}$Nb has been measured to decay to $^{92}$Zr by electron capture with a half-life of $\sim35$\,Myr~~\cite{MAKINO:1977,Nethaway:1978zz,Heinitz:2022}. The long-lived property makes $^{92}\text{Nb}$ a promising cosmochronometer~\cite{Lugaro:2016zuf,Haba:2021PNAS}, although the astrophysical conclusions drawn from these analyses should be reevaluated in view of the findings in the present paper. 

The issue is traced to the rate fitting coefficient $a_0$; changing it from -4.260150e+00 to -4.260150e+01 puts the decay half-life for the $^{92}\text{Nb}$ to $^{92}\text{Mo}$ channel at its lower limit $\sim10^{11}$ yr~\cite{nucwallet}.  We implement this correction, which is physically equivalent to setting the rate to zero. An additional fix in \texttt{SkyNet} involved modifying the normalization of the neutrino spectrum. This was necessary to reproduce the correct luminosities for the \lq\lq pinched\rq\rq\ spectra used here.

In the results shown in this paper, we used the density profiles from the outflow calculation as an input to \texttt{SkyNet} and permitted it to calculate the temperature using its internal self-heating subroutine. As a consistency check, we also ran calculations where both the \textit{density and temperature} profiles were provided as inputs to \texttt{SkyNet}. 

\begin{figure}
    \centering
    \includegraphics[width=0.8\textwidth]{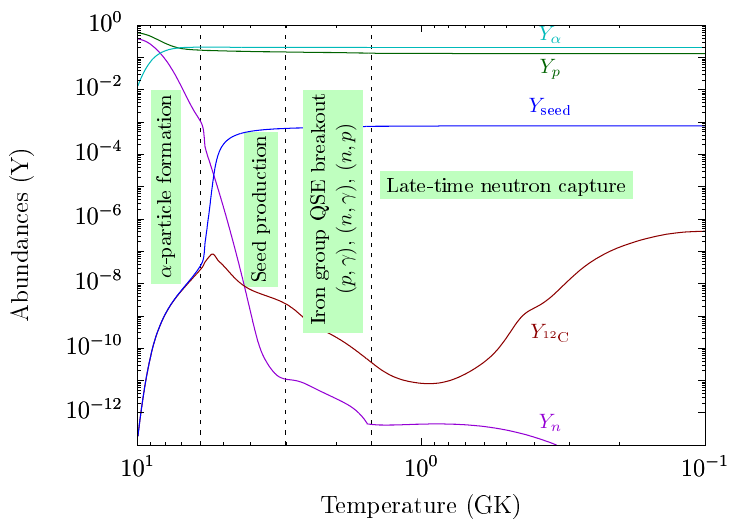}
    \caption{{Radial} evolution of the following nuclear abundances as a function of outflow temperature: $n$, $p$, $^{4}$He, $^{12}$C, and \lq seed\rq\ nuclei ($A \geq 12$), with temperature for a 1.4\,s post-shock outflow profile snapshot of the $13\,M_\odot$ progenitor model.}
    \label{fig:evol}
\end{figure}

Each nucleosynthesis run commences at a temperature of 10\,GK ($\simeq 0.86$\,MeV), in conditions of nuclear statistical equilibrium (NSE). \texttt{SkyNet} tracks the nuclear abundance $Y_{A,Z}$ of each species using an NSE calculation until a temperature of 9\,GK is reached, following which it switches to a full network calculation. Figure~\ref{fig:evol} depicts the radial evolution of certain selected abundances, plotted as a function of the outflow temperature, 
for a particular snapshot of the $13\,M_\odot$ progenitor model.

\bibliographystyle{unsrtnat}
\bibliography{Nucleo_bib}

\end{document}